\documentclass[review,12pts]{elsarticle}
\usepackage{epsfig}
\usepackage{pifont}
\usepackage{natbib}
\usepackage{geometry}
\usepackage{fleqn}
\usepackage{graphicx}
\usepackage{txfonts}
\usepackage{hyperref}
\usepackage{lineno}

\begin{document}

\title{TIARA: a large solid angle silicon array \\ for direct reaction studies with radioactive beams}

\author[dare,pais]{M.~Labiche\corref{cor}}
\ead{marc.labiche@stfc.ac.uk}

\author[surr]{W.N.~Catford}
\author[dare]{R.C.~Lemmon}
\author[surr]{C.N.~Timis}
\author[pais]{R.~Chapman}
\author[caen]{N.A.~Orr}
\author[liv]{B.~Fern\'andez-Dom\'inguez}
\author[pais]{G.~Moores}
\author[caen]{N.L.~Achouri}
\author[pais]{N.~Amzal}
\author[dare]{S.~Appleton}
\author[birm]{N.I.~Ashwood}
\author[surr]{T.D.~Baldwin}
\author[pais]{M.~Burns}
\author[val]{L.~Caballero}
\author[gan]{J.~Cacitti}
\author[gan]{J.M.~Casadjian}
\author[liv]{M.~Chartier}
\author[birm]{N.~Curtis}
\author[dare]{K.~Faiz}
\author[gan]{G.~de France}
\author[birm]{M.~Freer}
\author[caen]{J.M.~Gautier}
\author[surr]{W.~Gelletly}
\author[caen]{G.~Iltis}
\author[gan]{B.~Lecornu}
\author[pais]{X.~Liang}
\author[gan]{C.~Marry}
\author[caen]{Y.~Merrer}
\author[gan]{L.~Olivier}
\author[surr]{S.D.~Pain}
\author[dare]{V.F.E.~Pucknell}
\author[gan]{B.~Raine}
\author[gan]{M.~Rejmund}
\author[val]{B.~Rubio}
\author[gan]{F.~Saillant}
\author[gan]{H.~Savajols}
\author[gan]{O.~Sorlin}
\author[pais]{K.~Spohr}
\author[sac]{Ch.~Theisen}
\author[gan]{G.~Voltolini}
\author[dare]{D.D.~Warner}

\cortext[cor]{Corresponding author. Address: Daresbury Laboratory, Daresbury, Warrington, WA4 4AD, UK.}

\address[dare]{Nuclear Physics Group, STFC Daresbury Laboratory, Daresbury, Warrington, WA4 4AD, UK}
\address[pais]{Department of Engineering and Science, University of the West of Scotland, Paisley PA1 2BE, UK}
\address[surr]{Department of Physics, University of Surrey, Guildford GU2 5XH, UK}
\address[caen]{Laboratoire de Physique Corpusculaire, IN2P3-CNRS, ISMRA et Universit\'e de Caen, F-14050 Caen, France}
\address[liv]{Oliver Lodge Laboratory, University of Liverpool, Liverpool L69 7ZE, UK}
\address[birm]{School of Physics and Astronomy, University of Birmingham, Birmingham B15 2TT, UK}
\address[val]{Instituto de Fisica Corpuscular, CSIC-Universidad de Valencia, E-46071 Valencia, Spain }
\address[gan]{Grand Acc\'el\'erateur d'Ions Lourds, BP 55027,14076 Caen Cedex 5, France}
\address[sac]{Commissariat d'Energie Atomique de Saclay, 91191 Gif-sur-Yvette, France }

\begin{abstract}
A compact, quasi-4$\pi$ position sensitive silicon array, TIARA, designed to study direct 
reactions induced by radioactive beams
in inverse kinematics is described here. The Transfer and Inelastic All-angle Reaction Array (TIARA) consists 
of 8 resistive charge division detectors forming an octagonal barrel around the 
target and a set of double-sided silicon-strip annular detectors positioned at each end of the barrel.
The detector was coupled to the $\gamma$-ray array EXOGAM and the spectrometer VAMOS at the GANIL Laboratory 
to demonstrate
the potential of such an apparatus with radioactive beams. The 
$^{14}$N(d,p)$^{15}$N reaction, well known in direct kinematics, has 
been carried out in inverse kinematics for that purpose. 
The observation of the $^{15}$N ground state and excited states at 7.16 and 7.86 MeV is presented here as 
well as the comparison of the 
measured proton angular distributions with DWBA calculations. Transferred {\it l}-values are in very good 
agreement with both theoretical calculations 
and previous experimental results obtained in direct kinematics.    
\end{abstract}


\maketitle

\linenumbers

\section{Introduction}
A clear understanding of nuclear structure beyond the valley of $\beta$-stability 
requires detailed spectroscopic investigations.
Direct reactions, such as single-nucleon transfer reactions are established probes of the single-particle 
nuclear shell structure and have provided considerable insight into the properties of stable nuclei in the past. 
With the on-going increase in radioactive nuclear beam intensities, such as those achieved at the SPIRAL facility, this kind 
of reaction is now feasible. 
The inverse kinematics of such reactions leads, however, to significant constraints on the experimental 
apparatus \cite{Win97,Lenske98}. One of the main obstacles to overcome is to reach good energy resolution in 
the kinematically reconstructed excitation energy given that the energy spread
of the secondary beam may be relatively large, the target-like residue can be emitted over a large angular range and 
that thick targets are often required to compensate for the relatively low intensities of the beams \cite{Win97}. 
Already, pioneer detectors such as MUST \cite{Blum99} and the active target MAYA \cite{Demon07} have been build to tackle 
some of these obstacles and, the detector TIARA described here proposes a new alternative to these other apparatus. 
The TIARA array is designed and built specifically to study direct reactions with radioactive 
beams and addresses the challenge of the excitation 
energy resolution by employing the technique of $\gamma$-ray tagging. This has the advantage of providing, in principle, 
a final excitation energy resolution limited only by Doppler broadening.

The TIARA array was commissioned at the GANIL laboratory through a study of the d($^{14}$N,p)$^{15}$N reaction 
\cite{Phil69,Krets80} 
with coincident $\gamma$-ray detection. The results are reported here together with a full description of the array. 

\begin{figure}[ht]
\begin{center}
\includegraphics[width=10cm,height=7cm,angle=0]{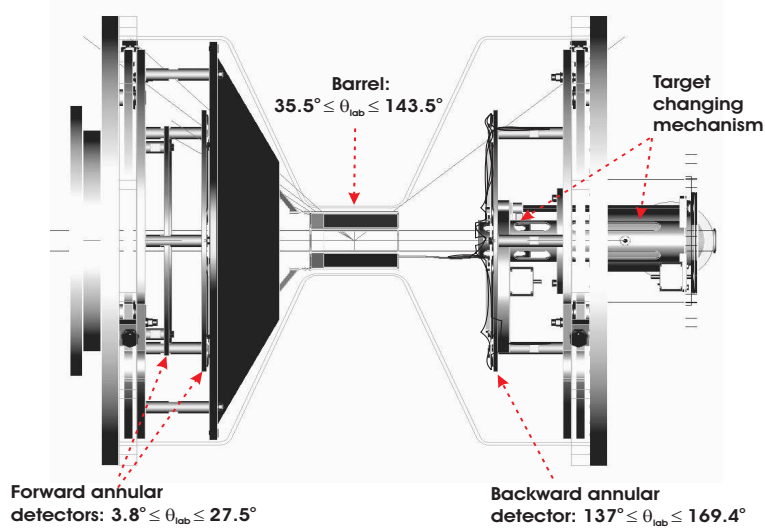}
\end{center}
\caption{Position of the silicon TIARA array and the target changing mechanism in the reaction chamber. The beam goes 
from the right to the left. 
The angular range covered by each component of the array for the commisioning experiment (see text) is shown.} 
\label{fig:TiaraCoverage}
\end{figure}

\begin{figure*}[ht]
\begin{center}
\includegraphics[width=10cm,height=6cm,angle=0]{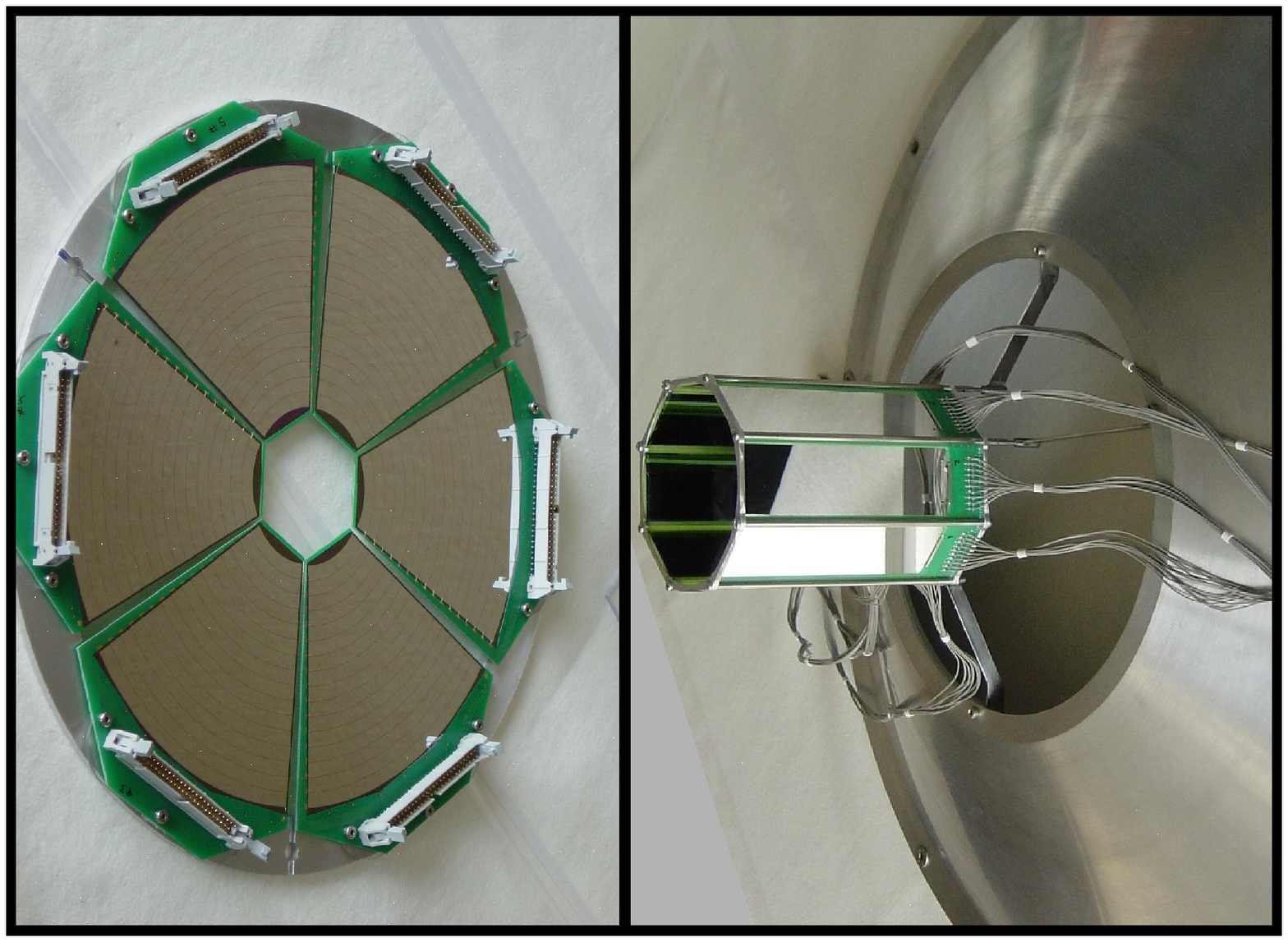}
\end{center}
\caption{The SiHyBall annular detector (left) and the octagonal barrel (right).} 
\label{fig:photoHybBar}
\end{figure*}

\section{Detector Description}
The TIARA array \cite{Cat02b,Cat04} has been designed with the ultimate goal of performing nucleon transfer 
and other direct reaction studies in 
inverse kinematics using radioactive ion beams \cite{Win97,Cat02}. 
The array is used to identify the binary reaction channels and to determine the excitation energies of the populated states.
This task is achieved by providing position and deposited energy measurements of the light charged target-like residue, which 
can be emitted over a wide angular range.
TIARA consists of a set of single-layer silicon detectors manufactured by Micron Semiconductor \cite{Micron} which covers 85$\%$ of 4$\pi$ 
(Fig.~\ref{fig:TiaraCoverage}). 
The set includes a large annular double sided silicon strip detector (SiHyBall), 
eight resistive charge division silicon detectors forming a ``barrel'' around the target, and two smaller ``CD-type'' silicon strip 
detectors (S1 and S2).

\subsection{Resistive Charge Division Detectors}
\begin{figure}[ht]
\begin{center}
\includegraphics[width=12cm,height=7cm,angle=0]{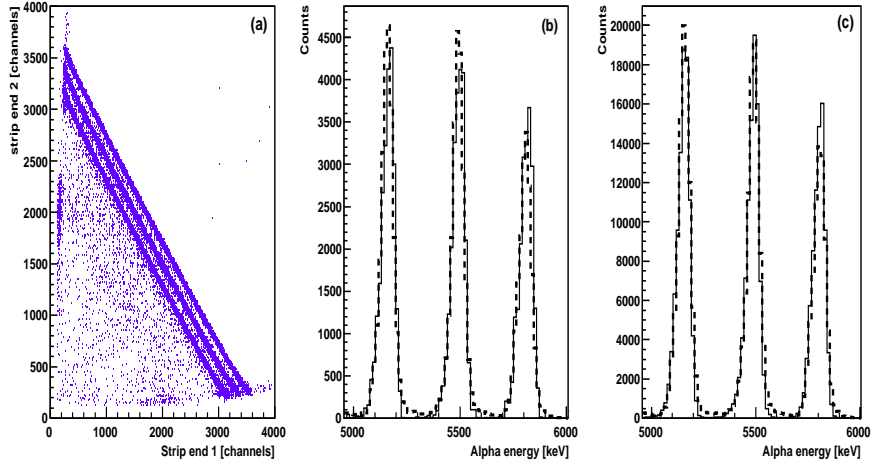}
\end{center}
\caption{Typical response of the barrel and SiHyball strips with a 3-alpha source. 
(a): signals collected at both ends of a barrel strip. 
(b): total energy collected in a single strip of a barrel detector (thick dashed histogram) and in a single strip of the 
DSS SiHyBall detector (thin line histogram), obtained with a 1$\mu$s shaping time. 
(c): same as (b) for the 4 strips of a barrel detector.}
\label{fig:3Alphas}
\end{figure}

Eight resistive charge division detectors based on 6-inch silicon wafer technology form an octagonal barrel around the beam 
axis and 
surrounding the target. Each of the detectors presents an active area 94.6 mm long and 22.5 mm wide with a thickness 
of 400 $\mu$m. 
The junction side facing the target 
is divided into 4 longitudinal resistive strips obtained by p$^{+}$ implantation on n-type silicon. 
Each 4k$\Omega$-resistive 
strip has a 5.65 mm pitch while the inter-strip gap is 100 $\mu$m. The strips provide for measurement and 
pixellation of the azimuthal angle in 32 bins of approximately 9.5$^{o}$.
The PCB board around the silicon has been minimised and bevelled so that the dead area between the detectors 
as well as between the barrel and 
annular detectors, is minimised. At one end of the Ohmic side of the detector, the PCB board is extended by $\sim$15 mm in order 
to gather all the output signal tracks: the 8 position signals (2 signals per strip) and the connection of the Ohmic side to ground.   
Miniature Junkosha coaxial cables of 1 mm diameter and 30 cm length were chosen for their favourable vacuum properties to transmit the 
signals from the detector to the vessel feed-throughs. 
Once assembled (Fig.~\ref{fig:photoHybBar} right), the barrel presents an octagonal cross section of 27.6 mm side length and 33.3 mm inner radius. From the centre, 
the angular range spans 36$^{o}$ to 144$^{o}$. For the commissioning measurements described later, the centre of the
barrel was mounted 1 mm forward of the target position leading to an angular coverage of 35.5$^{o}$ to 
143.5$^{o}$. The measurement of the position along the strip is achieved by resistive charge division and, with alpha particles of 5.5 MeV, the position resolution 
along the longitudinal axis is determined to better than 0.5 mm (FWHM). The resulting polar angle is thus deduced with a precision better than 1$^{o}$.  
The energy of the particle is obtained simply by summing the signals from the two strip ends. Figure~\ref{fig:3Alphas} was obtained using a mixed source of 
$^{239}$Pu, $^{241}$Am and $^{244}$Cm with alpha energies of 5156, 5484 and 5805 keV respectively. It illustrates the correlation between the 
signals from the ends of the strips (Fig~\ref{fig:3Alphas}(a)). With a shaping time of 1$\mu$s the barrel suffers slightly from ballistic deficits which result in 
a non-linear dependence of the energy sum, measured at each end of a strip, as a function of the position. Nevertheless, this dependence is easily 
described with a second-order polynomial function and a corresponding corrective factor can be applied to the energy sum. 
The resolution for one strip is $\sim$70 keV (FWHM) for 5.5 MeV alphas (Fig~\ref{fig:3Alphas}(b)).

\subsection{Annular Silicon Detectors}

As noted earlier, in order to enhance the angular coverage of the TIARA array, double-sided DC annular silicon-strip detectors are mounted at both ends of the 
barrel. For these detectors the annular rings on the entrance face (junction side) were fabricated by p$^{+}$ implantation on n-type silicon.

The forward angles are covered by two 500 $\mu$m thick annular detectors based on 4-inch wafer technology. The smallest of the two 
(S2-design) was positioned 150 mm downstream of the target position covering the polar angular range [3.8$^{o}$,13.1$^{o}$]. The active area is delimited 
by a disk of 11 mm inner radius and 35 mm outer radius. The detector is divided into 48 rings of 0.5 mm pitch at the front (target side) 
and 16 azimuthal sectors at the back. However, for the present measurements, the number of channels to instrument was reduced by linking 
the rings in threes giving effectively 16 rings of 1.5 mm pitch.

The second forward annular detector (S1 design) was mounted 92 mm downstream of the target position to cover the polar angular range [12.6
$^{o}$,27.5$^{o}$]. Its active area is divided into 4 quadrants of 20.5 mm inner and 48 mm outer radii. Although each quadrant 
has 16 front rings (1.65 mm pitch) and 4 azimuthal back sectors, for the experiment reported here the four quadrants were combined 
to form two semi-circles to reduce the total number of rings from 16$\times$4 to 16$\times$2.

The backward angles from 137.0$^{o}$ to 169.4$^{o}$ are covered by a 400 $\mu$m thick double-sided silicon-strip detector (DSSSD) based on 6-inch wafer technology and
positioned  150 mm upstream of the target position. This detector is composed of six individual wedges (Fig.~\ref{fig:photoHybBar}) originally 
developed at Oak Ridge for the SiHyBall forward array \cite{HyBall}. Each wedge is divided into 16 strips facing the target and 8 azimuthal back 
sectors. The active area of a wedge is delimited by inner and outer radii of 28.11 mm and 140 mm, respectively, and spans approximately 
55$^{o}$ of the total azimuthal angle. The pitch of the rings is 5.3 mm and the polar angular range is close to 2 degrees per strip. The 
energy resolution, illustrated in Fig.~\ref{fig:3Alphas}b, is typically $\sim$70 keV (FWHM) for 5.5 MeV alpha particles.

\begin{figure}[ht]
\begin{center}
\includegraphics[width=6cm,height=6cm,angle=0]{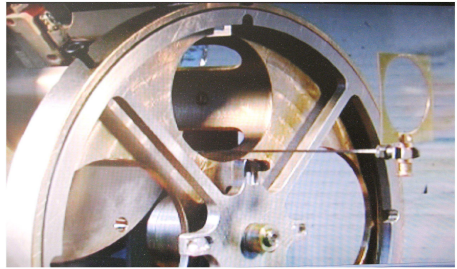}
\end{center}
\caption{The target mechanism on a test bench. The rod has just picked up a target frame from the storage wheel.} 
\label{fig:target}
\end{figure}

\subsection{Target Changing Mechanism}
One of the critical features of the TIARA array is the target changing mechanism (Fig.~\ref{fig:target}). The design of this mechanism
 has been chosen to maximise the solid angle coverage of the array. Positioned upstream, just behind the SiHyBall, it offers the possibility to use four different 
targets during a run without breaking vacuum. The mechanism consists of a target storage wheel with 4 positions 
and a rod parallel and slightly offset to the beam axis. A set of clamps, four on the storage wheel and one at the extremity of the rod (Fig.~\ref{fig:target}) 
are used to hold the target frames. The rod is driven along the beam axis via a ball screw. It first picks up a target from the wheel and continues 
its motion along the beam axis through the inner hole of the SiHyBall detector until the target position in the barrel is reached. 
The target frame is 3$\times$3 cm$^2$ in area with a central hole of 20 mm diameter. It can only be positioned perpendicular to the beam axis, 
introducing some shadowing at 90$^{o}$ in the barrel detector (Section 3.3). 
The whole mechanism is controlled remotely and the position of both the 
wheel and the rod is monitored by optical readouts. Four feed-throughs on the vacuum vessel are used for the target control system.

\begin{figure*}[ht]
\begin{center}
\includegraphics[width=14cm,height=10cm,angle=0]{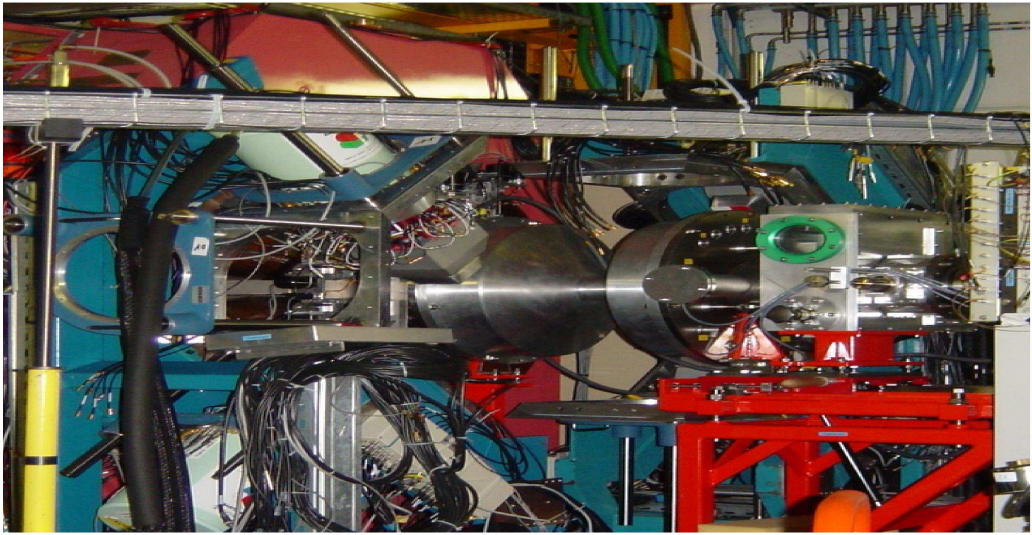}
\end{center}
\caption{Left: Picture of TIARA in situ. The support structure holding 4 EXOGAM Ge clover detectors has been opened up, showing the 
TIARA reaction chamber at the entrance of the VAMOS spectrometer. Right: The TIARA array and chamber as defined in GEANT4 simulation.} 
\label{fig:photoSetup}
\end{figure*}

\subsection{The Vacuum Vessel}
The reaction chamber of TIARA is made of aluminium and is some 56 cm long (excluding the target mechanism). Figure~\ref{fig:photoSetup} shows the vessel 
in position in front of the VAMOS spectrometer and in the middle of the EXOGAM support structure. The vessel presents a 
longitudinal diabolo shape with a central cylindrical section of 85 mm outer diameter housing the barrel and two 500 mm diameter cylindrical sections 
at each end housing the annular detectors.  
Two aluminium end plates accommodating Fischer DBPE 105-series feed-throughs (27 pins each) and supporting kinematics plates for detector alignment 
complete the chamber. While one of the end-plates can accommodate up to 17 feed-throughs, the other one, which also includes two pipes for additional
pumping, can accommodate up to 15 of them. Given that 4 feed-throughs are already used for the target mechanism, a total of 28 feed-throughs can be 
used for the transmission of the detector signals. 
The TIARA reaction chamber has been designed to allow a gamma-ray array such as EXOGAM to be placed as close as possible to the target. As such, the 
thickness of the walls of the central section has been limited to 2 mm in order to reduce the $\gamma$-ray attenuation to a minimum. For a photon energy 
of 1 MeV, the linear attenuation coefficient in Aluminium is 0.166 cm$^{-1}$. This leads to an attenuation of 3.3$\%$ in a 0.2 cm layer compare to 8$\%$ 
in a 0.5 cm layer. 

\subsection{Electronics and Data Acquisition}
 There are 8$\times$2$\times$4 channels to be instrumented for the octagonal 
barrel, 
(16 rings + 8 sectors)$\times$6 channels for the SiHyBall detector, 
(16 rings + 8 sectors)$\times$2 channels for the S1 detector and (16 rings + 16 sectors) channels for the S2 detector, resulting in a total of 288 channels. 
\begin{figure*}[ht]
\begin{center}
\includegraphics[width=10cm,height=10cm,angle=-90]{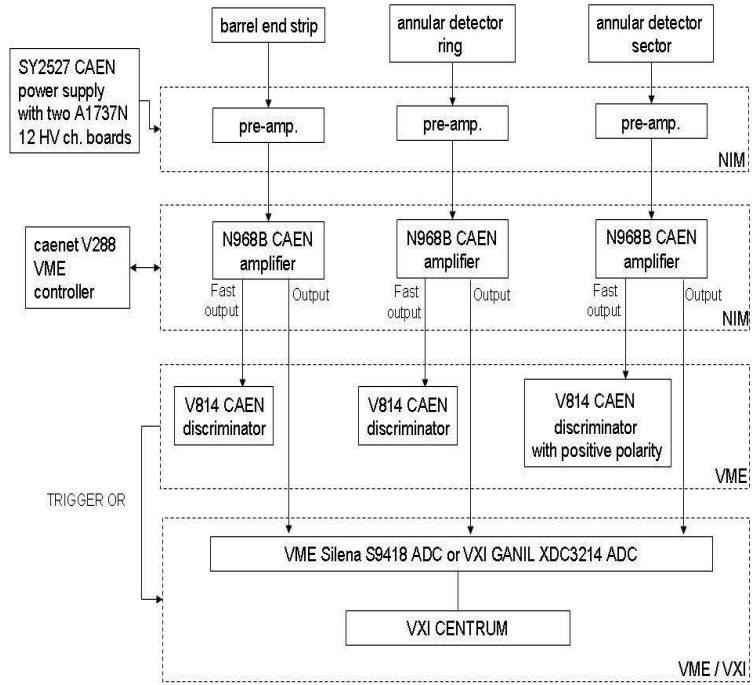}
\end{center}
\caption{Diagram of the TIARA electronics for one end of a barrel strip, one ring and one sector of the annular detectors (SiHyball, S1, S2).} 
\label{fig:TiaELEC}
\end{figure*}
A schematic diagram of the TIARA electronics is shown in Fig.~\ref{fig:TiaELEC}. 
Eighteen 16-channel charge-sensitive preamplifier modules manufactured at the University of the West of 
Scotland\footnote{Previously University of Paisley.}, eighteen CAEN N568B 16-channel spectroscopy amplifiers controlled remotely via a CAENET V288 
controller module, eighteen CAEN V814 16-channel 
low threshold discriminators and nine 32-channel ADC modules are employed to record the energy signals from the array. A SY2527 CAEN universal multi-channel power supply system equipped with a A1737N 
12 High Voltage (HV) channel board provided the -50 Volts necessary for the full depletion of all the silicon detectors. Typically leakage currents of around 0.2, 
0.3, 4.0 and 1.1 $\mu$A respectively are drawn by each element of the barrel and each wedge of the SiHyBall, the S1 and the S2 detectors.
   
The 16 charge-sensitive preamplifiers are mounted in double width NIM modules and are designed 
specifically for use with room temperature silicon-strip detectors and resistive-sheet detectors with capacitances in the range 
of 0 to 1000pF. Each unit houses two 8 channel motherboards 
with easily dismountable preamplifier chips. Both the motherboards and preamplifiers chips are housed in a rugged well shielded metal
housing.
With a quiescent DC output approaching zero, this unit is well adapted for use with the CAEN N568B spectroscopy amplifier, which has 50 
$\Omega$ input impedance.
For each of the six wedges of the SiHyBall detector, one complete module was used for the 16 front rings (with all 16 HV inputs combined) 
while half of another module was used for the eight back sectors. In this way, three preamplifier modules 
instrumented two wedges of the SiHyBall. Similarly three and two modules instrumented the S1 and S2 detector, respectively. 
For the four resistive strips of each of the eight barrel detectors, only half a module was required with all the corresponding eight 
preamplifier HV inputs combined and connected. 

Among the eighteen discriminators, five had to be adapted by the manufacturer to run with positive polarity inputs in order to instrument 
the eighty back sectors of the SiHyBall, S1 and S2 DSSSD annular detectors.
Both the CAEN amplifiers and discriminators are remotely programmable and, for TIARA, the control of this hardware is ensured via 
the Multi Instance Data Acquisition System (MIDAS)\cite{Puck} application developed at the STFC Daresbury Laboratory. Also used for the present 
work, and programmable via MIDAS, are the eight 32-channel GANIL XDC3214 ADCs \cite{GANILADC} operating in common dead-time mode and an additional 
32-channel Silena S9418 ADC.
        
EXOGAM, VAMOS and the TIARA array have their own stand-alone electronics and data acquisition systems (DAQs). For the TIARA commissioning 
measurements discussed below, the 3 DAQs were merged together using 3 hardware VXI CENTRUM modules which provided time stamping of the events and the
MERGER software for building the events \cite{Witt05}. The principal trigger of this commissioning experiment was defined by a hit in any element of TIARA. 

Due to room constraint around the TIARA vacuum chamber, a $\sim$3 m cable length was necessary to connect the TIARA detectors to the preamplifiers. As a direct 
consequence, the thresholds in energy had to be set high. They were $\sim$1 MeV for the double sided annular detectors and $\sim$1.5 MeV for the resistive 
charge division detector.         

\begin{figure*}[ht]
\begin{center}
\includegraphics[width=14cm,height=8cm,angle=0]{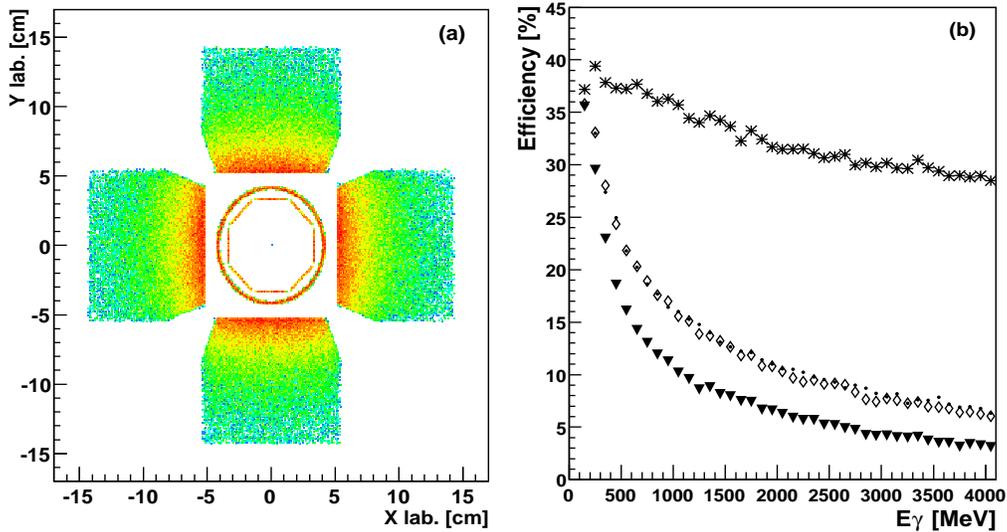}
\end{center}
\caption{ (a) Image of the TIARA and EXOGAM arrays reconstructed from the first interaction point of a simulated 1 MeV $\gamma$-ray. 
This is a projection in the plane perpendicular to the beam axis ($\it{z}$) and conditioned by -4 cm $\le$ $\it{z}$ $\le$ 4 cm. 
(b) Simulation of EXOGAM efficiency as a function of the $\gamma$-ray energy. The 4 clovers, denuded of their BGO Compton shields, are only 
~5 cm away from the target/source position. The triangles and the stars represent the photopeak and total efficiencies, respectively. The dots and 
diamonds represent the photopeak efficiencies after incorporating the ``addback'' procedure with and without the Lorentz boost.} 
\label{fig:TiaEXOSim}
\end{figure*}

\section{Commissioning}
\subsection{Experimental Details}

For the commissioning of the array and in order to validate the technique of heavy-ion---particle---$\gamma$ coincidence measurements, the 
d($^{14}$N,p$\gamma$)$^{15}$N reaction was investigated at an energy representative of SPIRAL radioactive beams. The 10.6 MeV/nucleon $^{14}$N beam was delivered by the first cyclotron of the GANIL facility.
The target was a 1 mg/cm$^{2}$ deuterated polythene (CD$^{2}$)$_{n}$ self-supporting foil.
The choice of the reaction was dictated by a number of considerations:
(a) the ground state of $^{15}$N is isolated and easily resolved from the excited states;
(b) the excited states populated in the reaction cannot be easily resolved in inverse kinematics and
(c) the reaction has been studied before in direct kinematics with light-particle detection at a similar centre-of-mass energy. 

 As shown in Fig.~\ref{fig:photoSetup} the TIARA reaction chamber allows four clovers of the EXOGAM array \cite{Sim00} to be mounted in a 
cube-like configuration. In this configuration the segmented Ge clover detectors are all positioned at 90$^{o}$ relative to the beam axis 
and from a distance between the target and the front face of each detector of approximately 5 cm. 
The photopeak efficiency in this configuration (Figure~\ref{fig:TiaEXOSim}) is 13.5$\%$ 
at 1.332 MeV when the 4 central contact signals of the 4 crystals in each clover detector are added together (``addback'').

The TIARA and EXOGAM arrays were mounted at the entrance of the VAMOS spectrometer (Fig.~\ref{fig:photoSetup}) operating 
in momentum-dispersive mode \cite{Savaj99}. The forward focused beam-like fragments are then identified in mass and charge 
from measurements of the time-of-flight, energy loss, residual energy and position in VAMOS.
A plastic finger was placed in front of the VAMOS focal plane detection system to intercept the intense non-interacting direct 
beam, which could damage the focal plane detectors. The data presented here were recorded over a total of 
approximately 4 hours of beam time with an average beam intensity of 2$\cdot$10$^{6}$ pps.

\subsection{Simulations and Data Analysis}
Knowing the efficiency of the experimental setup is essential if reaction cross sections and, hence, spectroscopic factors, are to be extracted.
In this context a complete and realistic simulation of the setup can be extremely useful.
 A Monte-Carlo simulation based on the GEANT4 code \cite{Agos02} has been developed to mimic the response of the TIARA and EXOGAM arrays. 
The geometry defined in this simulation includes, in particular, the entire active area of the TIARA array and the 2 mm thick aluminium walls of  
the reaction chamber, the four EXOGAM Ge clover detectors, and the target \cite{Labi05}. 
Figure~\ref{fig:TiaEXOSim}(a) illustrates a reconstructed image of the response of the setup to a 1MeV $\gamma$-ray.

For the simulation of nucleon-transfer reactions, the event generator takes into account the kinematics of the 2-body reactions and the 
differential angular cross section is set to be isotropic. The position of the proton source (or interaction) 
in the target is chosen randomly according to the beam spot size and the target thickness. The $\gamma$-rays 
are simulated assuming isotropic emission in the rest frame of the beam-like reaction product and then boosted by the Lorentz effect. 
The intrinsic resolutions of all the detectors are also included.

\begin{figure}[ht]
\begin{center}
\includegraphics[width=10cm,height=8cm,angle=0]{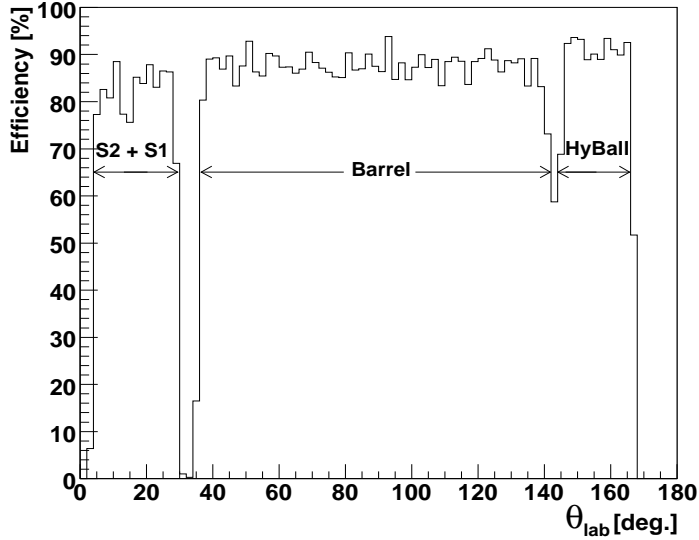}
\end{center}
\caption{Efficiency of the TIARA array as a function of the laboratory angle according to a GEANT4 simulation in which 2-5 MeV protons 
were generated and emitted isotropically.} 
\label{fig:EffAng}
\end{figure}

Taking into account the inactive regions of the Si detectors, the simulated overall efficiency of the TIARA array for proton detection with 
energies of a few MeV emitted isotropically was found to be 84$\%$. The efficiency of the various components of the array as a function of the 
polar angle is illustrated in Figure~\ref{fig:EffAng}. 

An isotropic $\gamma$-source at the target position and with variable energy was also simulated to estimate the EXOGAM photopeak efficiency and the 
result is shown in Figure~\ref{fig:TiaEXOSim}(b). The photopeak attenuation induced by the presence of the TIARA detectors and the reaction chamber
is about 5$\%$ at 1.332 MeV, with the silicon layer accounting for about 1$\%$.

The output of the simulation is recorded in a ROOT tree which includes as many leaves as channels for the two arrays. 
The simulated data and the real calibrated data can then be analysed identically using the same analysis code performing 
the ``add-back'' and Doppler corrections.
  
\begin{figure}[ht]
\begin{center}
\includegraphics[width=8cm,height=12cm,angle=0]{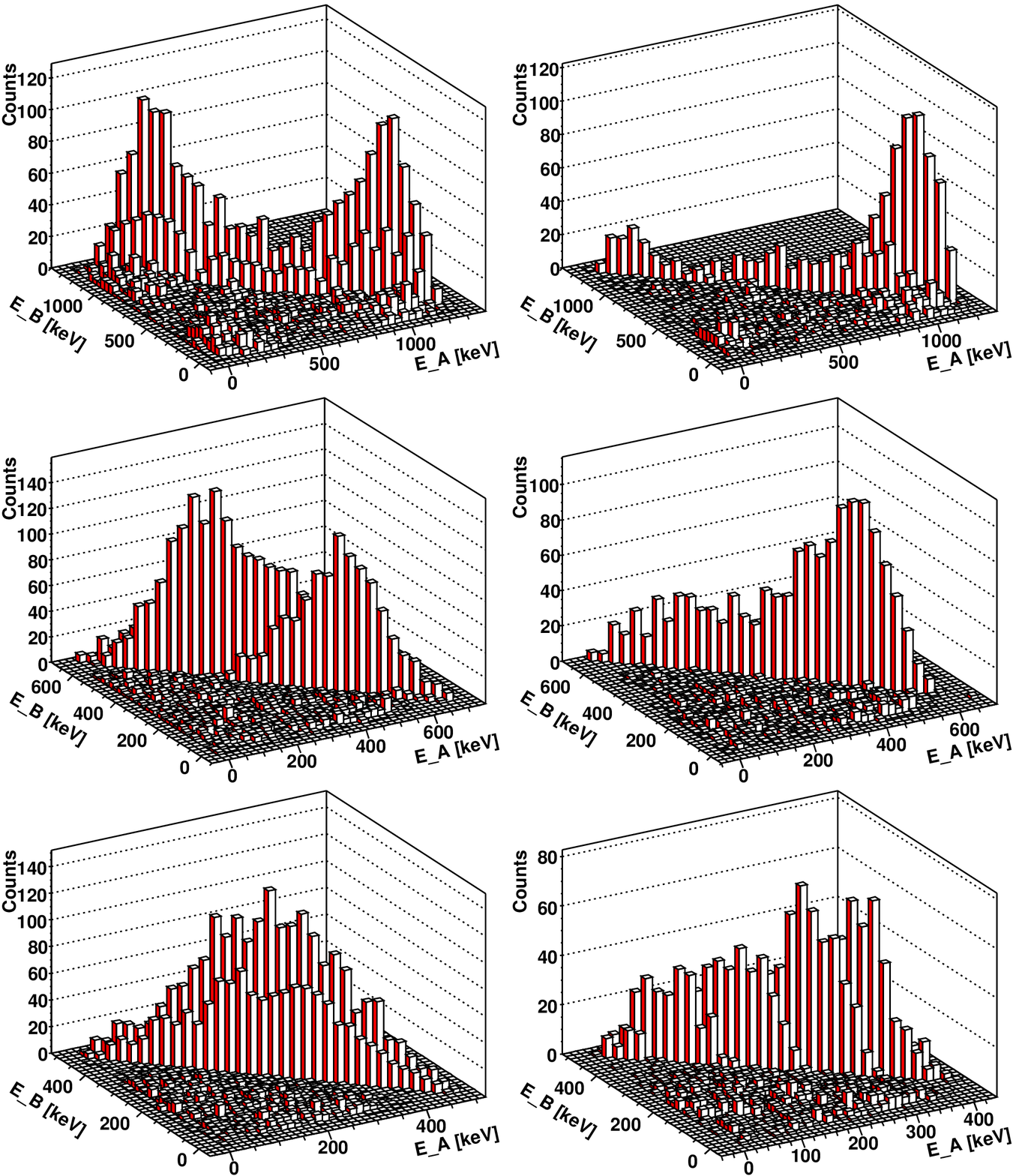}
\end{center}
\caption{ Top panels: Simulation of the energy deposited in crystal A versus the energy deposited in crystal B for an incident $\gamma$-ray of E$_{\gamma}$=1332 
keV, with no condition (left) and with the condition that crystal A is hit first (right). 
Middle Panels: E$_{\gamma}$=700 keV. 
Bottom panels: E$_{\gamma}$=500 keV.
} 
\label{fig:EXOGSim}
\end{figure}

An ``addback'' correction between the clover detectors was not considered here and, as noted above, was only applied to the 4 crystals of each clover. 
When more than one crystal was hit in a clover, the energies collected by the central contacts were summed together.  
The crystal with the highest deposited energy was taken to be, for the Doppler correction, the first crystal hit. 
Simulations (Figure~\ref{fig:EXOGSim}) show that this assumption is a valid approximation as long as the energy of the $\gamma$-rays 
is higher than 500 keV. Indeed, below 500 keV, when two crystals are hit the energies deposited in each crystal tend 
to be similar (see bottom-left panel of Fig.~\ref{fig:EXOGSim} ) and, as a consequence, the identification of the first crystal to be hit 
becomes uncertain.
For events of crystal multiplicity $M_{crys}=1$, the average angles chosen for the Doppler correction are 78${^o}$ for downstream crystals and 
102${^o}$ for upstream crystals.
For events of higher crystal multiplicity, the angles become respectively 84${^o}$ and 96${^o}$ as the closer to an
adjacent crystal a Compton interaction occurs the higher is the probability for $M_{crys}>1$. These angles have been determined empirically 
by matching of the photopeaks in downstream and upstream crystals and are consistent with the angles returned by the 
simulations.  

\subsection{Results}
\begin{figure}[ht]
\begin{center}
\includegraphics[width=9cm,height=9cm,angle=0]{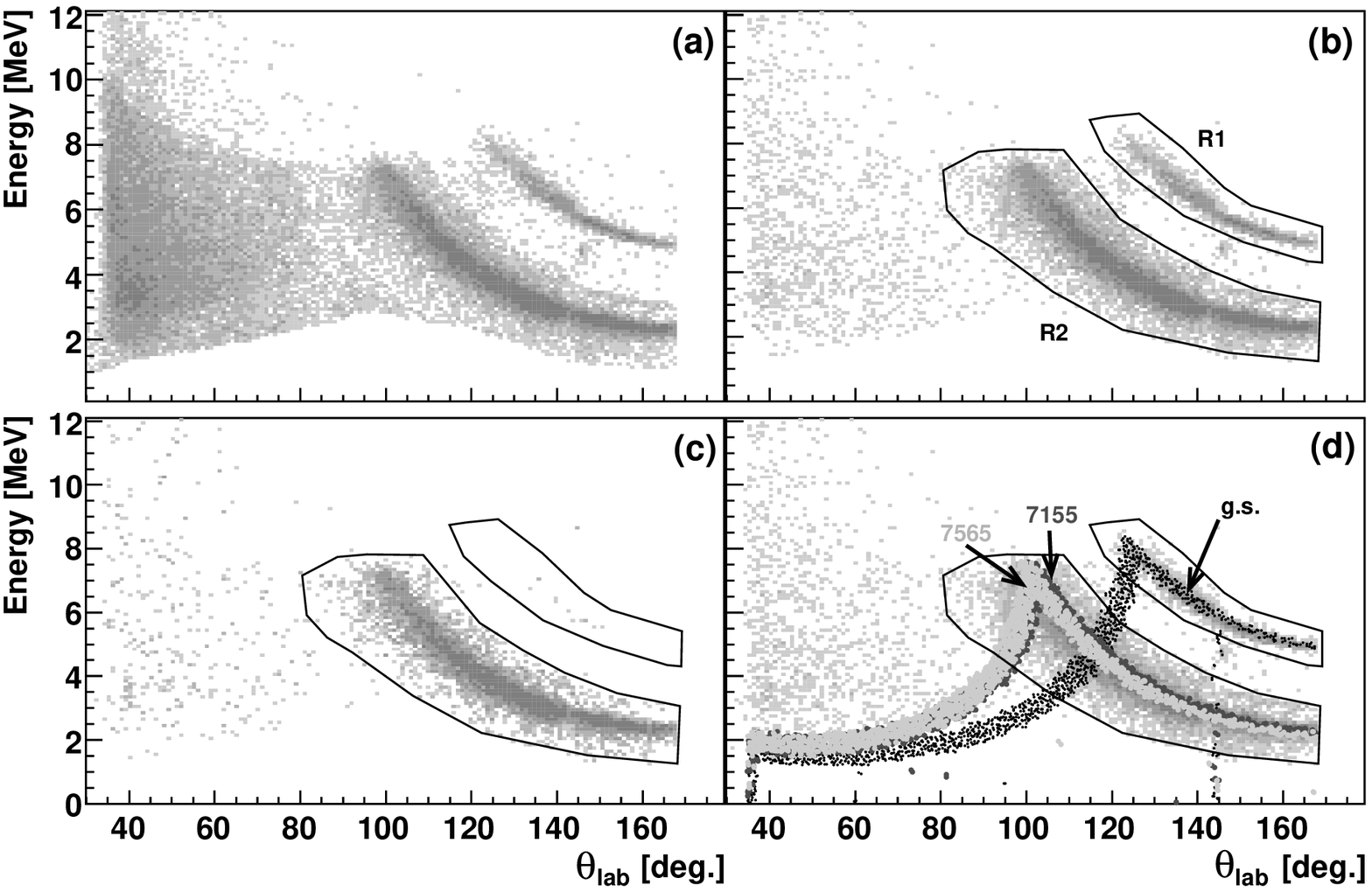}
\end{center}
\caption{ (a) Proton energy-angle spectrum for events detected in TIARA in coincidence with VAMOS. (b) Same as (a) with a gate 
on $^{15}$N identified in VAMOS. (c) Same as (b) with a coincidence in EXOGAM. (d) Same as (b) with 
Monte-Carlo simulations superimposed.} 
\label{fig:dE_Pos}
\end{figure}

The energy of the charged particles deposited in TIARA resulting from the reaction of the $^{14}$N beam with the (CD$_{2}$)$_{n}$ 
target as a function of laboratory polar angle recorded in the TIARA array is displayed in Figure~\ref{fig:dE_Pos}. The shadowing 
introduced by the presence of the target frame at 90$^{o}$ is noticeable.  
At backward laboratory angles where the emission of protons is expected, two clear kinematic loci are observed. These loci 
become even more pronounced when the $^{15}$N residue is identified in coincidence in 
the focal plane of the VAMOS spectrometer (Fig.~\ref{fig:dE_Pos}b). In the barrel detector, data associated with a low discriminator 
threshold have been removed, resulting in a noticeable inverted V-shaped cut at low deposited energy in Figs~\ref{fig:dE_Pos}a and b. 
The additional requirement of the detection of any $\gamma$-ray in coincidence, shown in Figure~\ref{fig:dE_Pos}c, leads to the disappearance of the 
protons in region R1 and, consequently, allows one to definitively associate this locus with the d($^{14}$N,p)$^{15}$N$_{gs}$ 
reaction. Indeed, when the kinematics of this reaction channel are used as input to the simulation, a perfect match between the simulation 
and the data (Fig.~\ref{fig:dE_Pos}d) is obtained. Both the kinematics and the expected proton punch-through energies are well reproduced. 
Note that the results of the simulations extend to very forward angles because the differential cross sections d$\sigma$/d$\Omega$ have, as noted earlier, been assumed to be isotropic. Since the reaction cross-section decreases relatively sharply with decreasing proton laboratory angle, the protons 
punching through the detectors are not so apparent in the data.

\begin{figure}[ht]
\begin{center}
\includegraphics[width=8cm,height=9cm,angle=0]{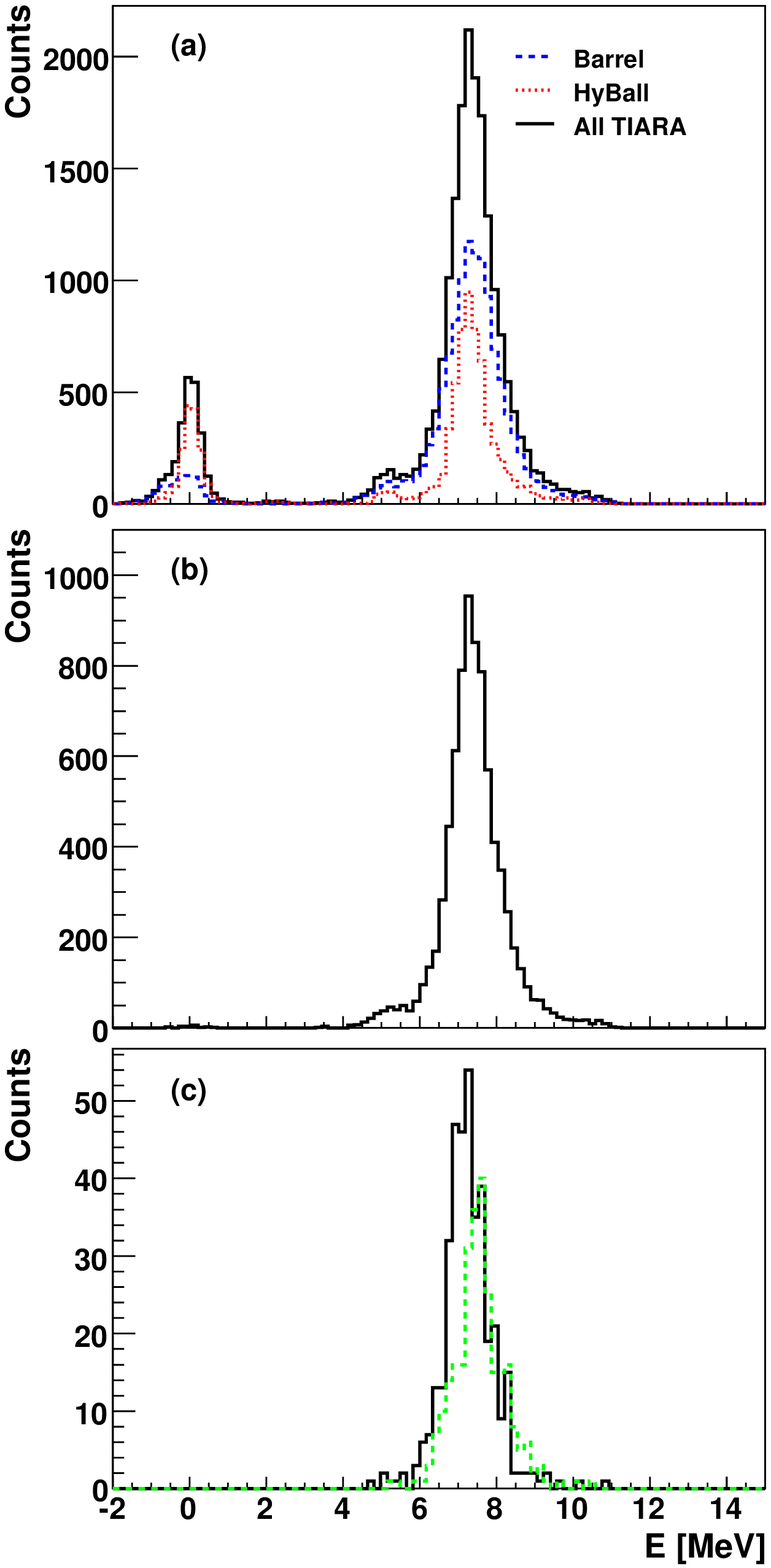}
\end{center}
\caption{ Excitation energy spectra obtained from the charge particles detected in TIARA: 
(a) With $^{15}$N detected in coincidence in VAMOS; 
(b) As for (a) with a $\gamma$-ray in coincidence in EXOGAM; 
(c) As for (a) with a 1885 keV (solid) and  2295 keV (dash) $\gamma$-ray in coincidence.}
\label{fig:Ex}
\end{figure}

Similarly, the simulation indicates that the protons observed in region R2 can be associated with the population of the 5/2$^{+}$ (7.16 MeV), 
3/2$^{+}$ (7.30 MeV) or 7/2$^{+}$ (7.57 MeV) states of the $^{15}$N.
The energy resolution of the excitation energy spectrum (Fig.~\ref{fig:Ex}) reconstructed from the proton energy and position 
measured in the barrel detector is  $\sim$1 MeV (FWHM) and is clearly insufficient to resolve the three states that lie above 7.1 MeV 
within 500 keV of each other. 
Apart from a broad structure centred at 7.5 MeV, Figure ~\ref{fig:Ex}(a and b) only reveals a small structure above 5 MeV. 
This can easily be interpreted as the direct population of the known 5/2$^{+}$ (5.27 MeV) and 1/2$^{+}$ (5.3 MeV) 
states. In addition, with a spectroscopic factor only 6 times smaller than the spectroscopic factors of the 7.16 and 7.57 MeV states, and 15 
times higher than the 5.3 MeV state\cite{Krets80}, the 5/2$^{+}$ state at 5.27 MeV is most probably the main contribution.     
         
\begin{figure}[ht]
\begin{center}
\includegraphics[width=9.5cm,height=12cm,angle=0]{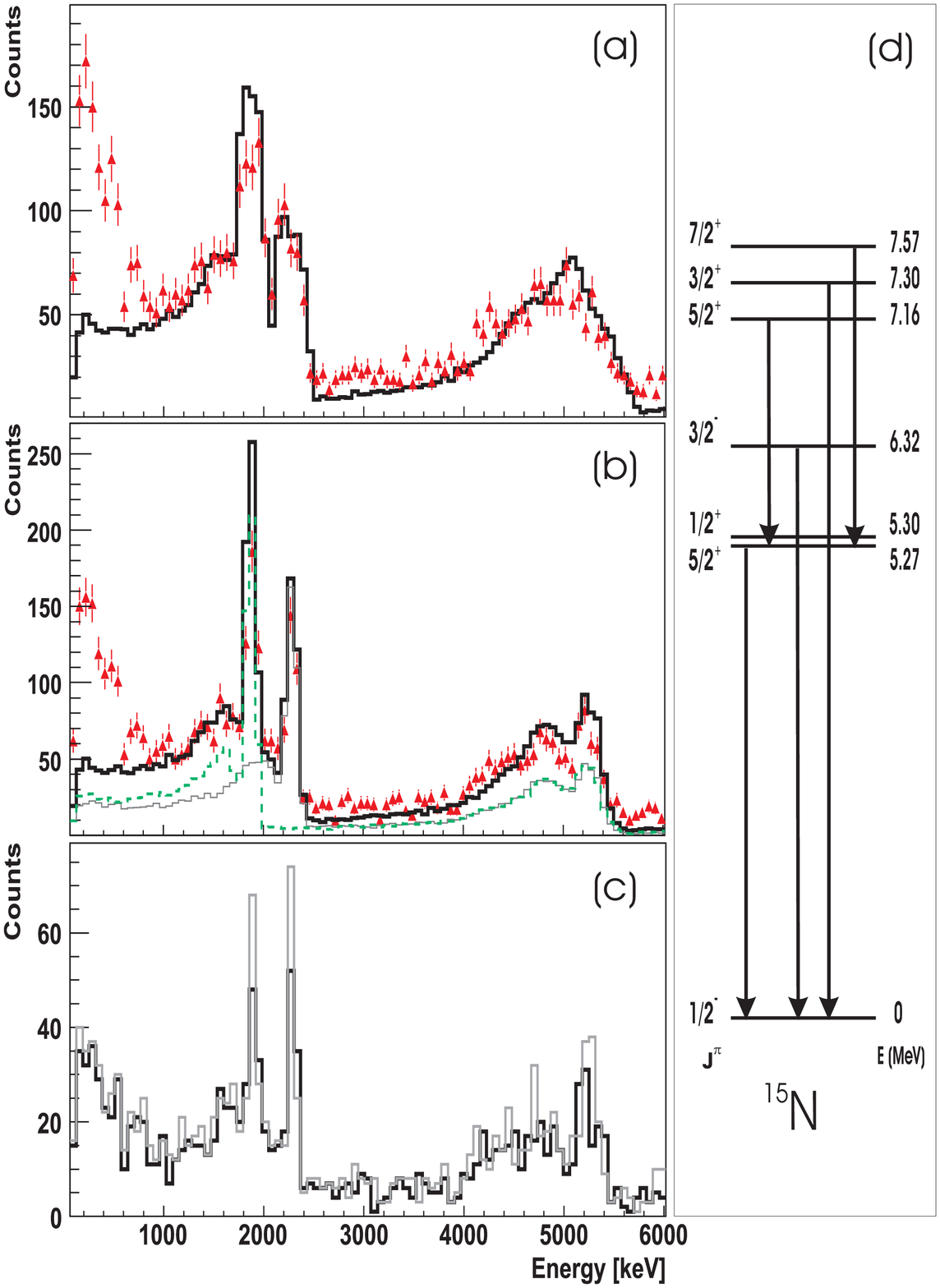}
\end{center}
\caption{ The $\gamma$-ray energy spectra in coincidence with the protons in region R2 of Fig.~\ref{fig:dE_Pos}. The data and the
simulations (histograms) of the $\gamma$-cascades from 5/2$^{+}$ and 7/2$^{+}$ levels to the ground state are shown before (a) and after (b) Doppler
correction.  The contributions of the cascades from the  5/2$^{+}$ and 7/2$^{+}$ levels at 7.16 and 7.57 MeV excitation energy are shown separately 
in (b). 
The distributions are normalized to the data by the integral of the number of events between 1 and 6 MeV. 
(c) Displays the data collected by a single clover after Doppler correction based on the 4 central contact signals (black histogram) and on 
the 16 outer contact signals (grey histogram). (d) Low energy $^{15}$N level scheme including the strongest transitions.}
\label{fig:ExogamData}
\end{figure}

The spectrum shown in Figure~\ref{fig:ExogamData} illustrates the crucial role played by 
the $\gamma$-ray array. All the spectra of Figure~\ref{fig:ExogamData} are conditioned by a clover multiplicity equal to 
one. The middle panel shows the energy distribution of the $\gamma$-ray measured in coincidence 
with the protons of region R2, after ``addback'' and Doppler correction. Two narrow peaks at 1885 and 2295 keV and a broader structure at 5270 keV 
which result from the de-excitation of the 5/2$^{+}$ and 7/2$^{+}$ states of $^{15}$N at 7.16 and 7.57 MeV, are observed (Figs.~\ref{fig:ExogamData}). 
As discussed earlier, the de-excitation of the 5/2$^{+}$ and 1/2$^{+}$ states at 5.27 and 5.30 MeV, populated directly in the reaction (Fig.~\ref{fig:Ex}),
also contributes to the broad structure at 5170 keV, to a small extent. 
It should be noted that the 3/2$^{+}$ level at 7.30 MeV, also observed by \cite{Phil69}, decays directly to the ground state (Fig.\ref{fig:ExogamData}d) 
and will, therefore, not be seen in coincidence with $\gamma$-rays in EXOGAM (owing to the very low detection efficiency at such high energies).
The Compton edges of the 1885 and 5270 keV $\gamma$-rays are also evident at around 1650 and 5000 keV respectively.
A simulation of the two decay cascades, taking into account the Lorentz boost and assuming that the two states were equally populated,
has been carried out. The result is displayed by the histogram in Figure~\ref{fig:ExogamData}(a) which has been normalized to 
the data such that the integrals between 1 and 6 MeV are the same. Although the intensities of the photopeaks at 1885 and 2296 keV seem slightly 
over-estimated, there is a good overall agreement between the simulation and the data. The discrepancy below 1 MeV is 
believed to arise from $\gamma$-rays scattering in material surrounding the detectors that have not been included in the simulation. 
Indeed, in both the data and the simulation, only the events for which a single clover detector is hit, have been taken into account. When all multiplicities 
are required, the number of counts at low energy increases in both the data and the simulated spectrum.
Figure~\ref{fig:ExogamData}(b) also displays the contribution of each cascade and, in particular, the contribution of the 
Compton background in the region of the photopeaks. According to the simulation these Compton events represent 26$\%$ of the peak intensity at 1885 
keV and $\sim$3.5$\%$ at 5270 keV. Given that the $\gamma$-decay of the two states (assumed here to be equally populated) proceeds via the 5.27 MeV state, 
half of the Compton contribution from the line at 5270 keV is actually in coincidence with 
the unobserved 1885 keV $\gamma$-ray. Therefore the total background contribution to the 1885 peak is 27.8$\%$. The Compton background from 
the 5270 keV line to the 2296 keV peak has similarly been estimated to be 2.5$\%$.

While, with the proton detection only, the final resolution on the excitation energy is restricted to $\sim$1 MeV, the gamma tagging technique 
improves dramatically the resolution
 to $\sim$100 keV, allowing for the two closely spaced $^{15}$N excited states to be resolved.  
During the experiment, the segmentation information of the Ge crystals was only available for one of the four clover detectors. 
A comparison of the $\gamma$-ray energy spectra recorded in this clover with and without segmentation information is shown in Figure~\ref{fig:ExogamData}(c). 
At 2.3 MeV, the resolution (FWHM) is 80 and 120 keV with and without the segmentation information, respectively.

\begin{figure*}[ht]
\begin{center}
\includegraphics[width=13.9cm,height=6.cm,angle=0]{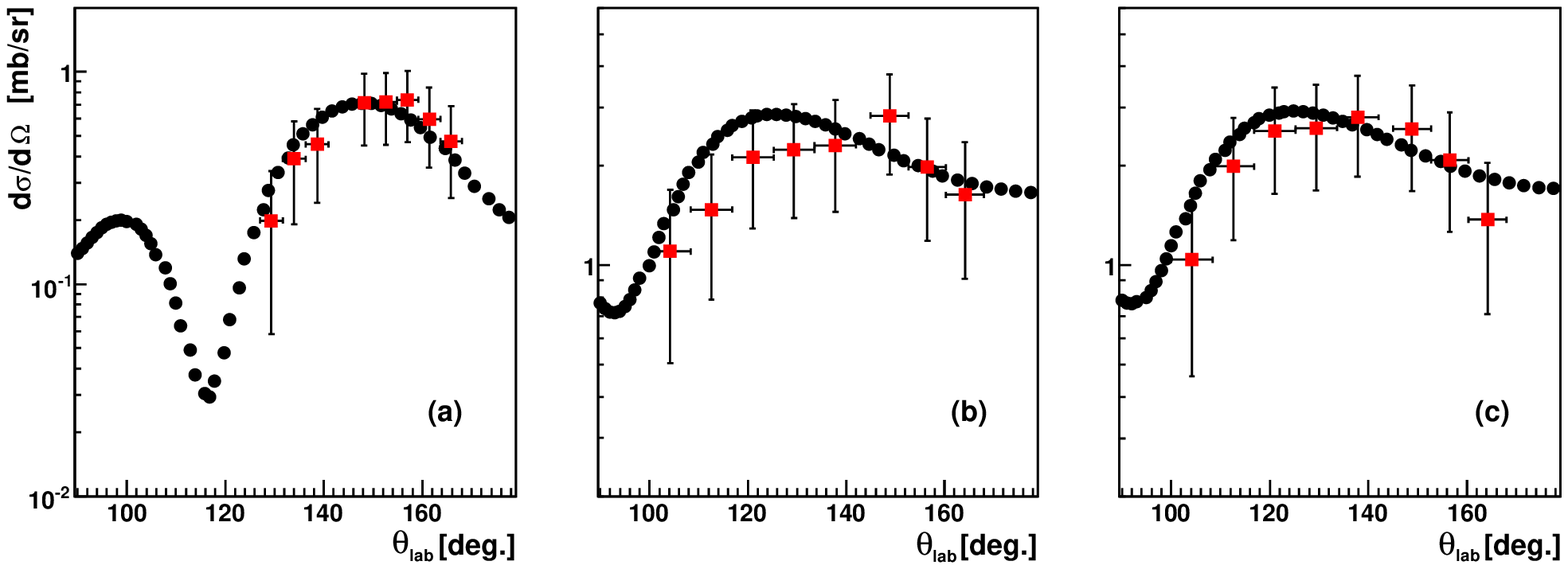}
\end{center}
\caption{ Proton differential cross section from the population of the ground (a) and excited states at 7155 (b) 
and 7565 keV (c) of $^{15}$N in the reaction d($^{14}$N,p). The filled circles show the DWBA calculations normalized by previously 
measured spectroscopic factors. The filled squares represent the data normalized to fit the DWBA calculations.
}  
\label{fig:DataDWBA}
\end{figure*}

The proton angular distributions displayed in Figure~\ref{fig:DataDWBA} were extracted by selecting events in region R1 of Figure~\ref{fig:dE_Pos}(b) 
corresponding to the $^{15}$N ground state, and in coincidence with the 1885 and 2296 keV $\gamma$-ray lines (Fig.~\ref{fig:ExogamData}b) corresponding 
to the $^{15}$N levels at 7.16 and 7.57 MeV.
  
The DWBA calculations displayed in Figure~\ref{fig:DataDWBA} were performed using the TWOFNR code \cite{Toyama} with the optical model parameters 
calculated according to the Johnson-Soper prescription \cite{Johnson70}. Each of the theoretical distributions have been multiplied 
by the corresponding spectroscopic factors derived from direct kinematics measurements in order to obtain the absolute differential 
cross section \cite{Krets80}: C$^{2}$S(g.s.)=1.33 , C$^{2}$S(7.16 MeV)=0.90  and C$^{2}$S(7.57 MeV)=0.88. 
Transferred angular momenta of {\it l}=1 for the ground state and {\it l}=2 for the two excited states were considered in the DWBA 
calculations in agreement with the results obtained in direct kinematics \cite{Phil69,Krets80}.
The experimental proton angular distributions for the ground and excited states were normalized to the theoretical 
distribution.
The shape of the experimental distributions for the ground state (Fig.~\ref{fig:DataDWBA}a) is in good agreement with the 
theoretical distributions. Similarly, the shapes of the experimental and theoretical distributions for the two excited states Figures 
~\ref{fig:DataDWBA}(b), and (c), are also in good agreement.

Unfortunately, in the present measurements, the absolute cross sections, and hence spectroscopic factors, could not be derived from the 
data. As noted earlier, part of the focal plane of VAMOS was protected from the transmitted beam by a ``finger''. As a result, no direct 
measurement of the beam dose or of the elastic scattering could be made.
Using, however, as a global normalisation the spectroscopic factor of 1.33 for the ground state \cite{Krets80}, the relative spectroscopic 
factors for the two excited states may be estimated. Values some 0.7 that of those previously measured \cite{Krets80}
were so deduced, which is within the statistical uncertainties of the measured differential cross section. 
Future measurements with radioactive beams will employ an active ``finger'' together with beam detectors also capable of counting the beam 
particles.

\subsection{Discussion}

In the present study, knowing the level scheme of $^{15}$N facilitates the identification of the populated states. For a nucleus with unknown 
level scheme, provided that there is enough statistics, a $\gamma$-$\gamma$ coincidence analysis can be carried out, in addition of simulations, 
to restore or establish a consistent level scheme.  Therefore, when coupled to a high efficiency $\gamma$-ray detector array, the TIARA array 
proposes a new alternative to other existing detectors for direct reaction with unstable beams. Other detectors like MUST offer a much higher 
dynamic range, better particle identification and intrinsic resolution than the TIARA array but they have a limited solid-angle coverage which 
can make (d,p) reactions studies difficult. On the other hand, active target detectors like MAYA competes very much with the TIARA array in term 
of solid-angle coverage and are known to have significant lower energy threshold than silicon detectors. However, the large volume occupied by 
active target detectors prohibits a coupling with a $\gamma$-ray array and, consequently, the $\gamma$-tagging technique can hardly be used.

\section{Summary}

A new compact, large solid-angle segmented silicon detector array, TIARA, designed for the study of direct reactions in inverse kinematics with 
radioactive beams, has been described. Coupled with a high efficiency $\gamma$-ray detector array, such as EXOGAM, TIARA employs the technique of light 
(target-like) particle-$\gamma$ coincidences to obtain the necessary resolution in excitation energy in the residual target-like recoil.
Identification of the latter, if required, may be performed using a magnetic spectrometer such as VAMOS. These techniques have been validated 
in a commissioning experiment in which the d($^{14}$N,p$\gamma$)$^{15}$N reaction was measured. In the near future, it is planned to increase the dynamic range of the silicon detector by the installation of a second 700 $\mu$m thick Si layer around the existing barrel detectors together with  15 mm thick CsI(Tl) segmented 
detectors backing the forward angle annular detectors. 

\section{Acknowledgements}

The collaboration wishes to thank the GANIL cyclotron operations crew for delivering the $^{14}$N beam. Partial support from the European Union under 
contract N$^{o}$506065 and from the Spanish MEC Grant FPA2005-03993 are also gratefully acknowledged. The development and construction of TIARA were financed by an EPSRC(UK) grant.

\end{document}